\documentclass[sigconf]{acmart}

\setcopyright{none}
\renewcommand\footnotetextcopyrightpermission[1]{}
\AtBeginDocument{%
  \providecommand\BibTeX{{%
    \normalfont B\kern-0.5em{\scshape i\kern-0.25em b}\kern-0.8em\TeX}}}

\acmConference[Conference acronym 'XX]{Make sure to enter the correct
  conference title from your rights confirmation emai}{June 03--05,
  2018}{Woodstock, NY}

\usepackage[export]{adjustbox}
\usepackage[ruled,vlined,linesnumbered]{algorithm2e}
\usepackage{amsmath,amsfonts}
\usepackage{appendix}

\usepackage{booktabs}
\usepackage{breakurl}
\usepackage{caption}
\usepackage{color}
\usepackage{enumitem}
\usepackage{fancyvrb}
\usepackage{filecontents}
\usepackage{graphicx}
\usepackage{hyphenat}
\usepackage{makecell}
\usepackage{marvosym}
\usepackage{multirow}
\usepackage{multicol}
\usepackage{listings}
\usepackage{pifont}
\usepackage{pythonhighlight}
\usepackage{subfigure}
\usepackage{tabularx}
\usepackage{tcolorbox}
\usepackage{textcomp}
\usepackage{tikz}
\usepackage{xcolor}
\usepackage{xspace}
\usepackage{threeparttable}
\usepackage{CJKutf8}
\usepackage{hyperref}

\newenvironment{code}{\captionsetup{type=listing}}{}

\begin{document}

\title{ConFL: Constraint-guided Fuzzing for Machine Learning Framework}



\author{ Zhao Liu} 
\affiliation{
    \institution{360 AI Security Lab}
    \country{China}}
    \email{liuzhao3@360.cn}

\author{ Quanchen Zou } 
\affiliation{
	\institution{360 AI Security Lab}
	\country{China}
}
\authornote{Corresponding author}
\email{zouquanchen@360.cn}

\author{ Tian Yu} 
\affiliation{
	\institution{360 AI Security Lab}
	\country{China}}
	\email{yutian@360.cn}

\author{ Xuan Wang} 
\affiliation{	
	\institution{360 AI Security Lab}
	\country{China}}
	\email{wangxuan3@360.cn}

\author{ Guozhu Meng} 
\affiliation{
	\institution{SKLOIS, Institute of Information Engineering, Chinese Academy of Sciences}
	\country{China}}
	\email{mengguozhu@iie.ac.cn}

\author{ Kai Chen} 
\affiliation{\institution{SKLOIS, Institute of Information Engineering, Chinese Academy of Sciences}
	\country{China} }
	\email{chenkai@iie.ac.cn}

\author{ Deyue Zhang} 
\affiliation{\institution{360 AI Security Lab}
	\country{China} }
	\email{zhangdeyue@360.cn}

\begin{abstract}
  As machine learning gains prominence in various sectors of society for automated decision-making, concerns have risen regarding potential vulnerabilities in machine learning (ML) frameworks. Nevertheless, testing these frameworks is a daunting task due to their intricate implementation. Previous research on fuzzing ML frameworks has struggled to effectively extract input constraints and generate valid inputs, leading to extended fuzzing durations for deep execution or revealing the target crash.

  In this paper, we propose ConFL, a constraint-guided fuzzer for ML frameworks. ConFL automatically extracting constraints from kernel codes without the need for  any prior knowledge. Guided by the constraints, ConFL is able to generate valid inputs that can pass the verification and explore deeper paths of kernel codes. In addition, we design a grouping technique to boost the fuzzing efficiency.
  
  To demonstrate the effectiveness of ConFL, we evaluated its performance mainly on Tensorflow. We find that ConFL is able to cover more code lines, and generate more valid inputs than state-of-the-art (SOTA) fuzzers. More importantly, ConFL found 84 previously unknown vulnerabilities in different versions of Tensorflow, all of which were assigned with new CVE ids, of which 3 were critical-severity and 13 were high-severity. We also extended ConFL to test PyTorch and Paddle, 7 vulnerabilities are found to date.
\end{abstract}

\ccsdesc[500]{Software and its engineering~Software testing and debugging; Software reliability}

\keywords{machine learning framework, operator collection, constraints extraction, constraint-guided fuzzing}

\maketitle

\section{Introduction}

Machine learning (ML) has transformed modern technology by offering efficient solutions for tasks such as image classification, speech recognition, and natural language processing. Alongside advanced algorithms, ML frameworks like TensorFlow, PyTorch, and Caffe serve as essential building blocks for machine learning services. These frameworks equip developers with comprehensive APIs for data processing, model training, and inference, thereby simplifying and expediting the creation of ML applications. As the most widely utilized machine learning framework, TensorFlow has been embraced by millions of developers and underpins machine learning systems at thousands of companies. This includes many of the world's largest machine learning users, such as Google, Apple, ByteDance, Netflix, Tencent, Twitter, and numerous others \cite{b43}. 

Despite their popularity, ML frameworks are not immune to common software vulnerabilities such as stack overflow, heap overflow, and memory corruption issues. For example, TensorFlow has had 432 CVE vulnerabilities to date. Such security problems can lead to the leakage of sensitive information, arbitrary code execution, and the potential compromise of ML systems. As the use of ML applications continues to increase, the risks associated with ML frameworks can be significantly amplified. Therefore, it is essential to identify vulnerabilities in ML frameworks to mitigate these risks.

However, finding vulnerabilities in ML frameworks is challenging due to their complex implementation. A typical ML framework consists of a \emph{frontend} that provides APIs for developers to ease model development and a  \emph{backend} that performs tasks such as matrix computation, model optimization, or hardware adaptation. \emph{Operators}, which enable communication between the frontend and backend, lie in the backend but can be invoked from the frontend. As the computation unit for ML frameworks, operators are the main target for vulnerability hunting. However, identifying these operators can be a laborious task. Furthermore, operators may have multiple parameters of arbitrary types and unclear constraints, increasing the difficulty of test input generation. Regular fuzzers, such as Peach~\cite{b19}, AFL~\cite{b20}, and libFuzzer~\cite{b34}, either require significant engineering efforts to translate input grammar or lack knowledge of input constraints, which makes them limited in testing ML frameworks.

Recently, a line of work has made progress in fuzzing ML frameworks. For instance, DocTer~\cite{b11} extract input constraints from API documentation and uses them to guide the test input generation for fuzzing machine learning API functions. FreeFuzz \cite{b12} executes collected code or models from open source with instrumentation to trace dynamic information for each covered operator, then leverages this information to perform fuzz testing. DeepRel~\cite{b40} builds on FreeFuzz to share mined valid inputs between similar functions. However, DocTer, FreeFuzz, and DeepRel partially or entirely depend on API documentation, which may not always be available or well-maintained. As a result, these approaches might not cover all functions in a library's APIs. Furthermore, not every function may be invoked in open-source code, highlighting the need for new input constraint inference techniques that do not rely on documentation or high-quality sample usages. Since API documentation can be incomplete, outdated, or inconsistent with code, the derived constraints may not be comprehensive enough, leading to less efficient testing. For instance, DocTer achieves only a 33\% valid input generation rate. IvySyn~\cite{b41} automatically identify DL kernel code implementations and adding fuzzing hooks to perform mutation-based fuzzing with type-aware mutations. Once a set of crashing kernels is obtained, IvySyn synthesizes high-level code snippets that can propagate the offending inputs through high-level APIs. However, IvySyn's approach of synthesizing code snippets may not always be effective in producing evidence of the vulnerability, especially if the code is complex or the vulnerability is deeply embedded in the system.

In this work, we introduce ConFL, an approach that addresses the limitations of previous methods by automatically extracting operator constraints from source code. We choose the Python frontend as the entry point to test operators in backend C/C++ kernel code. ConFL first traverses all the operators in the source code, collecting information such as operator name, operator call chain, and parameter names. Next, ConFL extracts constraints from the source code using static taint analysis, which can be categorized into four types: environmental constraints, dependency constraints, validation constraints, and logical constraints. ConFL then constructs two types of fuzzing templates using the operator information and constraints: data templates specify the shape, type, and value of an operator's parameters, while control templates determine the control flow of the operator. Guided by these constraints, ConFL generates high-quality, structurally and semantically valid test inputs to examine operators.

To demonstrate the effectiveness of our approach, we primarily evaluate its performance on TensorFlow. ConFL outperforms DocTer, FreeFuzz, DeepRel, and IvySyn in various aspects. ConFL demonstrates a higher code coverage, indicating its effectiveness in generating valid inputs and exploring a broader range of code paths. Furthermore, the success rate of ConFL is consistently higher, as it is able to execute more test cases without parameter errors or exceptions, ultimately leading to better vulnerability detection. Most notably, ConFL discovers 84 previously unknown vulnerabilities in different versions of TensorFlow, all of which have been assigned new CVE IDs, including 3 critical-severity and 13 high-severity vulnerabilities. We have also extended ConFL to test PyTorch and Paddle, uncovering 7 vulnerabilities to date. 

\noindent\textbf{Contributions.} We make the following contributions.
\begin{itemize}
\item Efficient operator collection and constraint extraction: ConFL effectively collects operators from machine learning frameworks, extracting environmental constraints, dependency constraints, validation constraints, and logical constraints to build comprehensive constraint trees.

\item Enhanced test data generation: Utilizing the extracted constraints, ConFL generates test input in a more guided and efficient manner, leading to a higher number of successful executions and improved code coverage compared to random generation or state-of-the-art fuzzers.

\item Increased vulnerability detection: By efficiently generating valid inputs, ConFL effectively identifies vulnerabilities in ML frameworks, enhancing the security and robustness of machine learning frameworks.

\end{itemize}

\section{Background \& Problem Statement}
\label{sec:problem}

\subsection{Typical Architecture of ML Framework}

An ML framework serves as a platform that simplifies the process of creating, training, and deploying machine learning models by providing pre-built libraries and tools for developers. The core functions of an ML framework, which involve mathematical algorithms for processing data and making predictions, are implemented in the kernels. Kernels are the central components of an ML framework that handle the low-level operations required for the framework, and developers access these functions through frontend interfaces.

For instance, TensorFlow, a popular ML framework, comprises a frontend and a backend. The frontend offers programming interfaces such as Python, Java, and C++, and constructs the computation graph. The backend, on the other hand, provides the runtime environment and executes the computation graph. It comprises four layers, namely the runtime layer, computation layer, network layer, and device layer. The runtime layer receives, constructs, and orchestrates the computation graph, while the computation layer offers kernel implementations of operators. The network layer implements inter-component communication, and the device layer supports various devices such as CPU, GPU, TPU, among others.

\begin{figure*}[htbp]
\setlength{\abovecaptionskip}{0.1cm}
\centering
\includegraphics[width=\textwidth]{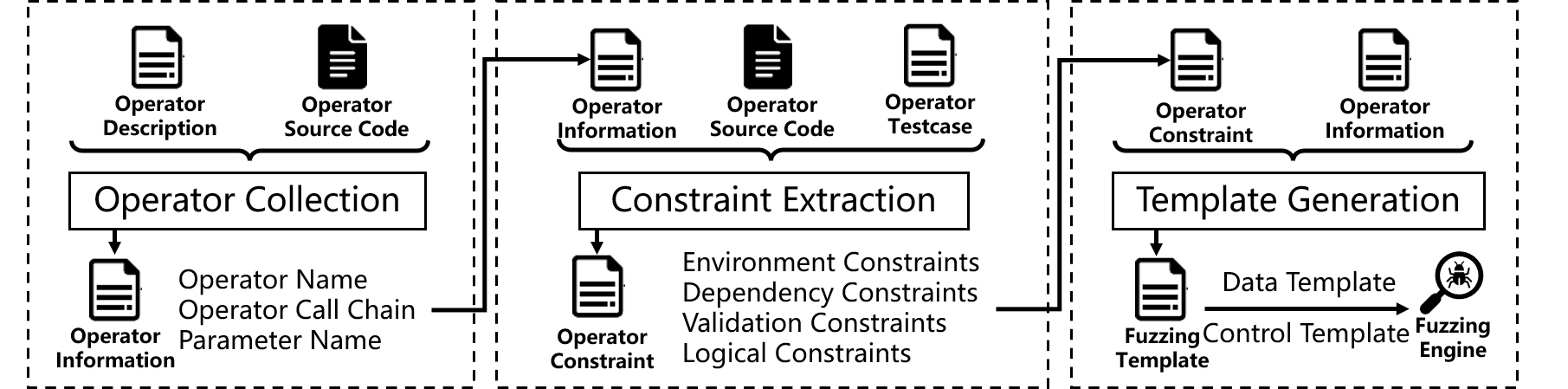} 
\captionsetup{font={footnotesize}}
\caption{Workflow of ConFL} 
\label{Fig.tool} 
\vspace{-0.4cm}
\end{figure*}

\color{black}\subsection{Operators of Machine Learning Libraries}

In this paper, our focus is on detecting vulnerabilities in operators used within machine learning frameworks. Operators serve as functions or operations that perform mathematical calculations on tensors or arrays of data, and are utilized to construct machine learning models. These operators form the building blocks of such models, and are responsible for performing tasks such as Conv3D for convolution or MaxPool for pooling. For efficient computation, operators are often developed and implemented in C/C++, while providing a Python interface to users.

We take the operator named \texttt{LoadAndRemapMatrix}(LARM)~\cite{b17} and \texttt{BoostedTreesCalculateBestFeatureSplit}(BTCBFS)~\cite{b27} provided in TensorFlow as examples. After analyze the description file in source code, We found that the operator LARM takes 9 parameters to load a tensor with name old\_tensor\_name from the checkpoint and BTCBFS takes 9 parameters to calculate gains for each feature and return the best possible split information for the feature.

\begin{table}[!ht]\small
    \setlength{\abovecaptionskip}{0.25cm}
    \setlength{\belowcaptionskip}{-0.cm}
    \vspace{0.25cm}
    \centering
    \captionsetup{font={small}}
    \caption{Illustrative examples of operators in Tensorflow} \label{bg_example}
    \begin{tabularx}{0.48\textwidth}{cX}
        \Xhline{1pt}
            \multicolumn{1}{c}{Operator} &
            \multicolumn{1}{c}{Parameter} \\
        \midrule 
            \multirow{2}{*}{LARM} & ckpt\_path,old\_tensor\_name,row\_remapping,col\_remapping,initializing\_values,  num\_rows, max\_rows\_in\_memory ,num\_cols ,name   \\  
            \multirow{2}{*}{BTCBFS} & node\_id\_range,stats\_summary,l1,l2,tree\_complexity, \\  & min\_node\_weight,logits\_dimension,split\_type,name  \\

        \Xhline{1pt}
    \end{tabularx}
    \vspace{-0.25cm}
\end{table}

\subsection{Problem Statement}

Fuzzing operators pose numerous challenges, making the process significantly more intricate than fuzzing conventional software systems. One key challenge is the complexity of operator functions, which include various tasks such as file management and computation. Additionally, implementations of operators differ across architectures (e.g., CPU, GPU), which further complicates testing against all operators. Another challenge lies in the complexity of input spaces for operators. They often deal with high-dimensional input spaces, like images or text sequences, making it difficult to generate meaningful and diverse inputs for fuzz testing within these spaces.

\textbf{Motivating Example.} Table 1 lists two operators that require multiple parameters, and these parameters have different types. The primary parameter type is called Tensor, which is a multi-dimensional array with a uniform type, such as int, float, or string.

We manually wrote the fuzzing templates to test these operators with random data. We spent a considerable amount of time collecting the data type of the operator parameters, which include float, int, char, string, and bool. We also considered the computing characteristics of the ML framework, including the list type and the Tensor type. An example of how to fill the BTCBFS operator with test data is given below:

\begin{code}
\setlength{\abovecaptionskip}{0cm}
\setlength{\belowcaptionskip}{0.15cm}
\begin{lstlisting}[language=Python]
tensorflow.raw_ops.BoostedTreesCalculateBestFeatureSplit(
    node_id_range=[1,7],
    stats_summary=[[[[2.0]], [[3.]], [[3.]]]],
    l1=[0.0],
    l2=[0.0],
    tree_complexity=[1.0],
    min_node_weight=[0.7],
    logits_dimension = 2,
    split_type = 'equality'
)
\end{lstlisting}
\captionsetup{font={footnotesize}}
\captionof{figure}{Cases for testing BTCBFS.}
\end{code}

However, upon running the BTCBFS test script, we encountered a type error message `\texttt{InvalidArgumentError: hessian dim should be < 0, got -1.}'

When testing BTCBFS, we narrowed the range of random data generation by analyzing the range of operator parameters in the document. For example, stats\_summary is four-dimensional, and logits\_dimension is an integer larger than 0. Although relatively normalized test data is generated, it still cannot be executed because valid data cannot be generated. Once the test input is invalid, the computing process will be terminated in the Python frontend, making it difficult to deeply test the specific code of the operator.At the same time, the word \texttt{hessian} in the error message is not in the operator parameter name list, which makes it more difficult to adjust the test data.

\subsection{System Overview}

In this paper, we focus on generating semantically valid test inputs for operators. Since extracting constraints from API documentation is incomplete and requires domain knowledge, we opt to automatically extract operator constraints from the source code. Our goal is to generate valid test inputs by leveraging constraints to  pass parameter validation detection successfully in the C++ backend. To achieve this, we have developed a prototype tool called ConFL, which consists of three modules, as shown in Figure \ref{Fig.tool}.

\textbf{Operator Collection.} ConFL aims to test operators, so the first step is to collect operator information. This module automatically traverses all the operators of an ML framework and collects information including operator name, operator call chain, and parameter names. Additionally, it can construct call chains to be used in fuzzing. This module is further explained in section 3.1.

\textbf{Constraints Extraction.} In this module, we extract constraints from the source code and categorize them into four categories: environmental constraints, dependency constraints, validation constraints, and logical constraints. Environmental and dependency constraints are used to restrict the execution context of operators to ensure they have access to the necessary resources for execution. Validation and logical constraints are used to generate valid and diverse test inputs to enable execution of the deeper code of the operator. We will provide a detailed description of this module in Section 3.2.

\textbf{Template Generation.} Using operator information and constraints, ConFL generates fuzzing templates, which provide an abstract representation of the operator before specific testing. There are two types of templates: data templates and control templates. Data templates specify the shape, type, and value of parameters, while control templates specify the control flow of the operator. By using the operator information, ConFL builds a skeleton for the template, and then relies on constraint information to build dependencies for different parameters. This module will be described in detail in section 3.3.
\section{Methodology}

\subsection{Operator Collection}

\textbf{Rationale for collecting operators in Python front-end code.}
As previously mentioned, ML frameworks like TensorFlow can be logically divided into two parts: the frontend, which provides interfaces in various programming languages for developers, and the C/C++ backend, which aims to enhance computing efficiency. We have selected the Python frontend as the entry point for testing the backend C/C++ code for several reasons:

\begin{enumerate}
[itemsep=1.3pt,topsep=0pt,parsep=0pt]
\item It is consistent with the practical scene. Most developers build neural networks for model training and inference with the Python frontend interfaces. 


\item Python offers excellent language features. Unlike IvySyn, which tests from the C++ side, ConFL chooses the Python side as the operator's input. The Python frontend has rich types, such as int, float, tensor, etc., which can guide the generation of valid parameters. Additionally, writing C/C++ harnesses for each operator is challenging, whereas generating operator templates automatically with Python's reflection mechanism saves a considerable amount of time.
\end{enumerate}


ML frameworks connect Python code with C/C++ code using pybind11, SWIG~\cite{b35}, and other methods, which are loaded into the Python runtime as modules. By selecting the Python frontend as the entry point, ConFL automatically traverses functions, classes, and modules with the help of Python's reflection mechanism, obtaining operator package directories. With operator names and package directories, ConFL analyzes the operator's signature and extracts parameter names, and generates an operator test template by concatenating this information.

\textbf{Algorithm.} Algorithm \ref{al_op_collect} demonstrates how ConFL collects operators and generates the test templates in detail. In the process of collecting operators, a tree structure of Module-Operator is constructed in line 1, which represents the call path of an operator from a leaf node to the root. ConFL collects modules in \texttt{getMods}. From line 5, we iteratively traverse all available modules. First, get modules from the parent module in line 6. Then, modules are added to the tree in lines 9-15 to obtain the full call path for each operator. Since there may be multiple call paths for an operator or a module, when a duplicate operator is detected, the one with the shortest call path is preserved, as shown in lines 10-12. Finally, return \texttt{modules} which contains all modules in the ML library. After collecting modules, ConFL gathers operators from modules with function \texttt{getOps}. It traverses \texttt{modules} from line 18, gets operators from each module in line 20, and adds operators to the tree in lines 22-27. Similar to \texttt{getMods}, duplicated operators are taken into consideration in lines 23-25. Eventually, after collecting all operators, traverse the Module-Operator tree and generate harness for each operator.

As a result, there are 9689 operators in total after the initial collection. Without any omission or manual writing, ConFL can automatically generate test templates for all interfaces in TensorFlow. 

\textbf{Further selection and deduplication.} ConFL automatically analyses the security of C/C++ backend through the Python frontend interface. Therefore, ConFL will remove the operators whose computation can be accomplished in the frontend. For example,  \texttt{tensorflow.experimental.dtensor.job\_name()} doesn't execute codes in C/C++ backend, so it will not be tested later.

With the Python's \texttt{id} function, we obtain unique interfaces identified by the memory addresses. However, we have observed that even when different Python interfaces possess distinct implementations, their corresponding C function call chains might be identical. To address this, ConFL further deduplicates operators by considering both the operator parameters and their call chain. As illustrated in Figure 3, the three interfaces shown share the same parameters, and the second and third interfaces have identical call chains. Consequently, we only generate test templates for the first two interfaces.

\begin{code}
\setlength{\abovecaptionskip}{0cm}
\setlength{\belowcaptionskip}{0.15cm}
\begin{lstlisting}[language=Python]
OP: tensorflow.reshape(tensor=[1,2],shape=[1,2])
Path: Dispatch -> TFE_Py_FastPathExecute -> TFE_Py_Execute

OP: tensorflow.raw_ops.Reshape(tensor=[1,2],shape=[1,2])
Path: TFE_Py_FastPathExecute -> TFE_Py_Execute

OP: tensorflow._compat.readers.array_ops.gen_array_ops.reshape(
    tensor=[1,2],shape=[1,2])
Path: TFE_Py_FastPathExecute -> TFE_Py_Execute
\end{lstlisting}
\captionsetup{font={footnotesize}}
\captionof{figure}{Function call chain in C++}
\end{code}

\textbf{Adaptility.} In the process of generating operator test templates, different ML framework frontends may have different implementation types. Taking the Python frontend as an example, there are functions or classes. For functions, the calling statements will be automatically constructed. For classes, an instance of the class will be generated first, and then the calling codes.

\begin{algorithm}[!ht]\small
    \caption{Operators Collection}
    \label{al_op_collect}
    \LinesNumbered
    \KwIn {A specific ML library: $mlLib$ }
    \KwOut {A Module-Operator Tree: $tree$}
    
    \SetKwFunction{FGetMods}{getMods}
    \SetKwFunction{FGetOps}{getOps}
    \SetKwProg{Fn}{Function}{:}{}
    tree.init(mlLib) 
    
    \Fn{\FGetMods{$parentMod$}}{ 
        modules = Queue() 
    
        modules.put(mlLib) 
        
        \While{modules not empty}{
            parentMod = modules.get()
            
            mods = getModMembers(parentMod)
            
            parentModPath = getModInfo(parentMod)
            
            \For{mod in mods}{
              \If{mod is duplicated}{
                \tcp{preserve the shortest call path when the same modules exists}
                \If{len(path(mod)) > len(path(parentModPath+mod.name))}{
                
                  tree.moveNode(mod, parentMod)
                  
                }
              }
              modules.put(mod) 
              
              tree.addNode(mod, parentMod)
            }
        }
        \Return{modules}
    }
    
    \Fn{\FGetOps{$modules$}}{ 
        \While{modules not empty}{
            parentMod = modules.get()
            
            ops = getOpMembers(parentMod)
            
            parentModPath = getModInfo(parentMod)
            
            \For{op in ops}{
              \If{op is duplicated}{
                \tcp{preserve the shortest call path when the same ops exists}
                \If{len(path(op)) > len(path(parentModPath+op.name))}{
                
                  tree.moveNode(op, parentMod)
                  
                }
                
            }
            tree.addNode(op, parentMod)
            }
        }
        \Return tree
    }

\end{algorithm}

\subsection{Operator Constraints Extraction}


The runtime behavior of an operator is dependent on the input data and the environment in which it is executed. ConFL meticulously extracts constraints within operators to ensure comprehensive coverage. We define the conditions necessary for an operator's successful execution as constraints and classify them into four types, including environmental constraints, dependency constraints, validation constraints and logical constraints. By thoroughly examining these constraints, ConFL achieves higher code coverage and uncovers vulnerabilities hidden in deep execution paths (described in detail in Section 4).

\subsubsection{Environmental Constraints}
In ML framework, there are various execution options available, and we refer to the constraints that determine the choice of execution mode as environmental constraints. For example, TensorFlow primarily offers two modes of executing operations: Eager Execution and Graph Execution. Eager execution is an imperative programming mode in which TensorFlow operations are executed immediately as they are called from Python. This mode is more intuitive and flexible, allowing for easier debugging and experimentation. With eager execution, users can work with TensorFlow operations just like any other Python operations, and there is no need to explicitly build a computational graph before executing it. Graph execution, also known as static computation graph, is the traditional mode of execution in TensorFlow. In this mode, users first define a computational graph that represents their model or algorithm, and then TensorFlow executes the graph in an optimized manner using a session. Graph execution offers performance benefits through various optimizations like parallelism, distributed execution, and efficient memory allocation. 

Since TensorFlow 2.0, eager execution is the default mode, but users can still use graph execution through the \texttt{tf.function} decorator, which converts user's Python code into a static graph. This allows users to leverage the benefits of graph optimizations while keeping the flexibility of eager execution.

Besides, Tensorflow use a domain-specific compiler named Accelerated Linear Algebra(XLA) for linear algebra to accelerate computaion.

\begin{table}[H]\footnotesize
    \centering
    \captionsetup{font={small}}
    \caption{Environmental Constraints} \label{env_cons}
    \begin{threeparttable}
    \begin{tabularx}{0.45\textwidth}{cll}
        \Xhline{1pt}
        \textbf{Category} & \textbf{Type} & \textbf{Constraints} \\ 

        \midrule
        \multirow{3}{*}{Execution Mode}
         & Eager execution & -  \\ 
         & Graph execution & @tf.function \\ 
         & XLA & @tf.function(jit\_compiler=True) \\ 
        \midrule
        \multirow{3}{*}{Architecture Mode}
         & CPU & - \\
         & GPU & with tf.device('/device:GPU:2') \\
         & TPU & TPUClusterResolver(tpu='') \\
        \Xhline{1pt}
    \end{tabularx}
    \vspace{0.1cm}
      
    \end{threeparttable}
\end{table}

\subsubsection{Dependency Constraints}
Dependency constraints refer to the parameter constraints that an operator must satisfy before actual execution. If the types and data of the parameters do not meet the operator's requirements, the execution of the operator will not commence. We divide dependency constraints into resource-dependent constraints and operation-dependent constraints.

\textbf{Resource-dependent constraints.} Various types of parameters are required during the computation process of ML framework operators. In TensorFlow, besides simple types such as int and float, there are also special types like resource and variant. The resource type represents a handle to a mutable, dynamically allocated resource, while the variant type represents data of an arbitrary type\cite{b39}. Based on the composition characteristics of operator parameters, we classify the types into two categories according to their complexity. All types in TensorFlow are shown in Table \ref{tableII}.

\begin{enumerate}[itemsep=1.3pt,topsep=0pt,parsep=0pt]
\item Basic types: Scalar like int, float, complex, char and string.
\item Composite types: Basic tensor, which is a combination of basic types. Resource tensor, such as a file handler or a series of codes.
\end{enumerate}

\begin{table}[!ht]\small
\vspace{-0.15cm}
\setlength{\abovecaptionskip}{0.2cm}
\setlength{\belowcaptionskip}{0cm}
    \centering
    \captionsetup{font={small}}
    \caption{Operator Types.} \label{tableII}
    \begin{tabularx}{0.45\textwidth}{c|l|l}
        \Xhline{1pt}
        \multirow{2}{*}{Basic Type} & 
        \multicolumn{2}{c}{Composite Type} \\
        \cline{2-3}
        & Basic Tensor & Resource Tensor \\
        \midrule
        bool & DT\_INT8/16/32/64 & DT\_RESOURCE \\ 
        int  & DT\_UINT8/16/32/64 & DT\_VARIANT \\
        float & DT\_BOOL & CODE \\
        string & DT\_COMPLEX64/128 & FILE \\
        char & DT\_QINT8/16/32 &  ~ \\
        ~  & DT\_QUINT8/16 & ~ \\
        ~  & DT\_HALF & ~ \\
        ~  & DT\_FLOAT & ~ \\ 
        ~  & DT\_DOUBLE & ~ \\
        ~  & DT\_BFLOAT16 & ~ \\
        ~  & DT\_STRING & ~ \\ 
        \Xhline{1pt}
    \end{tabularx}
    \vspace{0.1cm}
\end{table}


In the code repositories of ML frameworks like TensorFlow and Paddle\cite{b34}, operator description files are typically used to dynamically generate code at compile time or track historical changes in operator code. TensorFlow's operator description file is called \texttt{ops.pbtxt} (located in source code), which contains the operator name, parameter name, and type. By parsing the aforementioned parameter information, ConFL can obtain the types of all parameters. For example, \texttt{ckpt\_path} is a tensor of \texttt{DT\_STRING} type, and \texttt{num\_rows} is a tensor of \texttt{DT\_FLOAT} type. Additionally, the return value types can be extracted, such as the result of LoadAndRemapMatrix being a tensor of \texttt{DT\_FLOAT} type.




After parsing the type and value information of each parameter of LoadAndRemapMatrix, ConFL generates the following intermediate description:

\begin{code}
\setlength{\abovecaptionskip}{0cm}
\setlength{\belowcaptionskip}{0.15cm}
\begin{lstlisting}[language=Python]
    'ckpt_path': ['DT_STRING'],
    'old_tensor_name': ['DT_STRING'],
    'row_remapping': ['DT_INT64'],
    'col_remapping': ['DT_INT64'],
    'initializing_values': ['DT_FLOAT'],
    'num_rows': ['int'], 
    'num_cols': ['int'], 
    'max_rows_in_memory': ['int']
\end{lstlisting}
\captionsetup{font={footnotesize}}
\captionof{figure}{The parameters' type of LoadAndRemapMatrix.}
\end{code}




We find that there are dependencies between different operators, which means the output of one operator is used as the input of another operator. However, such construction of parameters is not reflected in the documentation or source code. We call this constraint as resource-dependent constraints. and we save the output type of the successfully executed operators. By analyzing the operator type, we can abstract the resource-dependent constraints to construct correct parameters.

Different operators may depend on various types of file data, making manual generation a labor-intensive task. For instance, LoadAndRemapMatrix requires loading a model file in ckpt format during the execution process. Since the input parameter data type is string, it represents the storage path of the model file. If the data of the string type is mutated, the operator cannot read the model file data during execution, resulting in execution failure. 

To address this issue, we propose to automatically extract relevant file pre-constraints with the assistance of test cases. TensorFlow contains an extensive collection of test cases. When testing a specific operator, the test case will include the code for the pre-deployment environment, such as generating the specified file. The code for the LoadAndRemapMatrix test is stored in checkpoint\_ops\_test.py, which contains the following code:

\begin{code}
\setlength{\abovecaptionskip}{0cm}
\setlength{\belowcaptionskip}{0.15cm}
\begin{lstlisting}[language=Python]
class LoadAndRemapMatrixTest(test.TestCase):
  def setUp(self):
    ...
    matrix = variable_scope.get_variable(
          'matrix',
          dtype=dtypes.float32,
          ...)
    save = saver.Saver([matrix])
    save.save(...)
\end{lstlisting}
\captionsetup{font={footnotesize}}
\captionof{figure}{LARM's testcase.}
\end{code}

By instrumenting the LoadAndRemapMatrix operator and monitoring the execution path of the drive letter, we can identify the corresponding file generated when the test case is executed.

\textbf{Operation-dependent constraints.} Operators are the smallest computing units in ML frameworks. According to our analysis, most operators have few calling dependencies, allowing for individual testing. However, some parameters may be of special types that require results generated by other operators as their inputs. ConFL identifies operators such as \texttt{pop}, \texttt{push}, and \texttt{close} through keyword matching, extracts the operator entity, and tests the operators with the same entity as a single group. For instance, Stack-related operators, like \texttt{StackPop}, \texttt{StackPush}, and \texttt{StackClose}, all share the Stack main body and construct relevant data for testing through built-in test sequences.

Different operators may operate on the same entity, such as Stack. If testing is performed only for a single operator, the operation dependency might not be satisfied. For example, the Push operation first requires initializing a Stack, while the Pop operation needs both the initialized stack as a parameter and data in the Stack, requiring the execution of the Push operation. We refer to the preceding operations that ensure the smooth execution of operators as operation-dependent constraints.

Operator names typically describe their functions semantically, such as StackClose, StackPush, and StackPop. By conducting part-of-speech analysis, we can identify the operations and entities within the operator name and consider different operators acting on the same entity as a set. In terms of operator execution sequence, if a test case detects that operators in the set are called in a specific order through the hook method, the relevant sequence is saved. If no relevant test exists in the test case, operation dependencies are determined through random execution.

During part-of-speech determination, since the position of a word affects the part-of-speech judgment, we shift the sequence after word segmentation to the left and save the verb part-of-speech tokens identified in all operators. After excluding the verb tokens, we assess the operator's name, and ultimately cluster the operators of the same subject.

\begin{code}
\setlength{\abovecaptionskip}{0cm}
\setlength{\belowcaptionskip}{0.15cm}
\begin{lstlisting}[language=Python]
LookupTableFind ['Find']
LookupTableRemove ['Remove']

ReaderReadUpTo ['Read']
ReaderRestoreState ['Restore']

Stack ['Stack']
StackClose ['Stack', 'Close']
StackPush ['Push']
\end{lstlisting}
\captionsetup{font={footnotesize}}
\captionof{figure}{The operator's name and operation.}
\end{code}

\subsubsection{Validation Constraints}




Environmental constraints and dependency constraints are mainly used to arrange the execution environment of the operator, so that the operator can have executable resources, but the execution conditions of the operator is also related to the constraints of input parameters. For example, the explicit shape, type and value constraints of every parameter in BTCBFS are obtained with the above methods. However, we find that there are dependencies between parameters. For example, the third value of \texttt{stats\_summary} in operator BTCBFS needs to be larger than \texttt{logits\_dimension}.

Validation constraints in operators refer to the parameter's conditions or rules that must be satisfied for the code to execute correctly and produce the expected output. These constraints play a crucial role in ensuring data integrity, maintaining API stability, and preventing errors or exceptions during the execution of a operator. If input parameters do not satisfy semantic rules, test cases often fail the semantic checks and falter in the shallow code of the operator. Consequently, only a small portion of inputs generated from generic generation-based fuzzing reaches the operator execution stage, where deep bugs typically hide, leaving a large part of the operator code unreached.





In this section, we propose a constraint extraction technique for operators. It can analyze the source code of operators, locate semantic checking statements, extract specific values compared with parameters, and ultimately perform as validation constraints. The validation constraints can be categorized into type constraints and numerical constraints, which serve as a guide for generating valid parameters in the subsequent stages of testing. The process consists of three main steps: first, compiling the source code into LLVM's\cite{b25} intermediate representation (IR) using clang\cite{b26}.; second, specifying taint sources, propagations, and sinks; and finally, extracting constraints at the taint sinks.

In Tensorflow, operators are implemented by extending OpKernel and overriding the compute method.  Operator parameters are divided into \texttt{input} and \texttt{attr}, as indicated by the \ding{172} symbol in Figure \ref{BTCBFS Taint Code}. \texttt{Inputs} are tensors with mutable values, while \texttt{attrs} remain constant from step to step. The operator receives the  \texttt{attr} parameter in the constructor, and the \texttt{input} parameter in the compute method. As a result, we select \texttt{context->GetAttr} (dotted box in Figure \ref{BTCBFS Taint Code}) and \texttt{context->input} (solid box in Figure \ref{BTCBFS Taint Code}) as sources. The first parameter of the function is the name of the Python positional parameter, while the second parameter represents the specific parameter name. We have identified seven types of source points:


\begin{code}
\setlength{\abovecaptionskip}{0cm}
\setlength{\belowcaptionskip}{0.15cm}
\begin{lstlisting}[language=C]
context->input(INDEX)
context->input("VARNAME", &VAR)
context->input_list("VARNAME", &VAR));
context->mutable_input(INDEX, _);
context->mutable_input(VARNAME, &VAR, _));
context->mutable_input_list("VARNAME", &VAR));
context->GetAttr("VARNAME", &VAR));
\end{lstlisting}
\captionsetup{font={footnotesize}}
\captionof{figure}{The operator's name and operation.}
\end{code}


As the operator primarily computes using input parameters, the return value of the function that retrieves inputs is designated as the taint source. Instructions such as \texttt{load}, \texttt{store}, and \texttt{getelementptr} act as the primary targets for taint propagation analysis. When a tainted variable is present in the operands of an instruction, the return variable of that instruction is marked as tainted. Taint propagations are denoted by \ding{173} in Figure \ref{BTCBFS Taint Code}.

\begin{figure}[H]
\setlength{\abovecaptionskip}{0.15cm}
\centering
\includegraphics[width=0.5\textwidth]{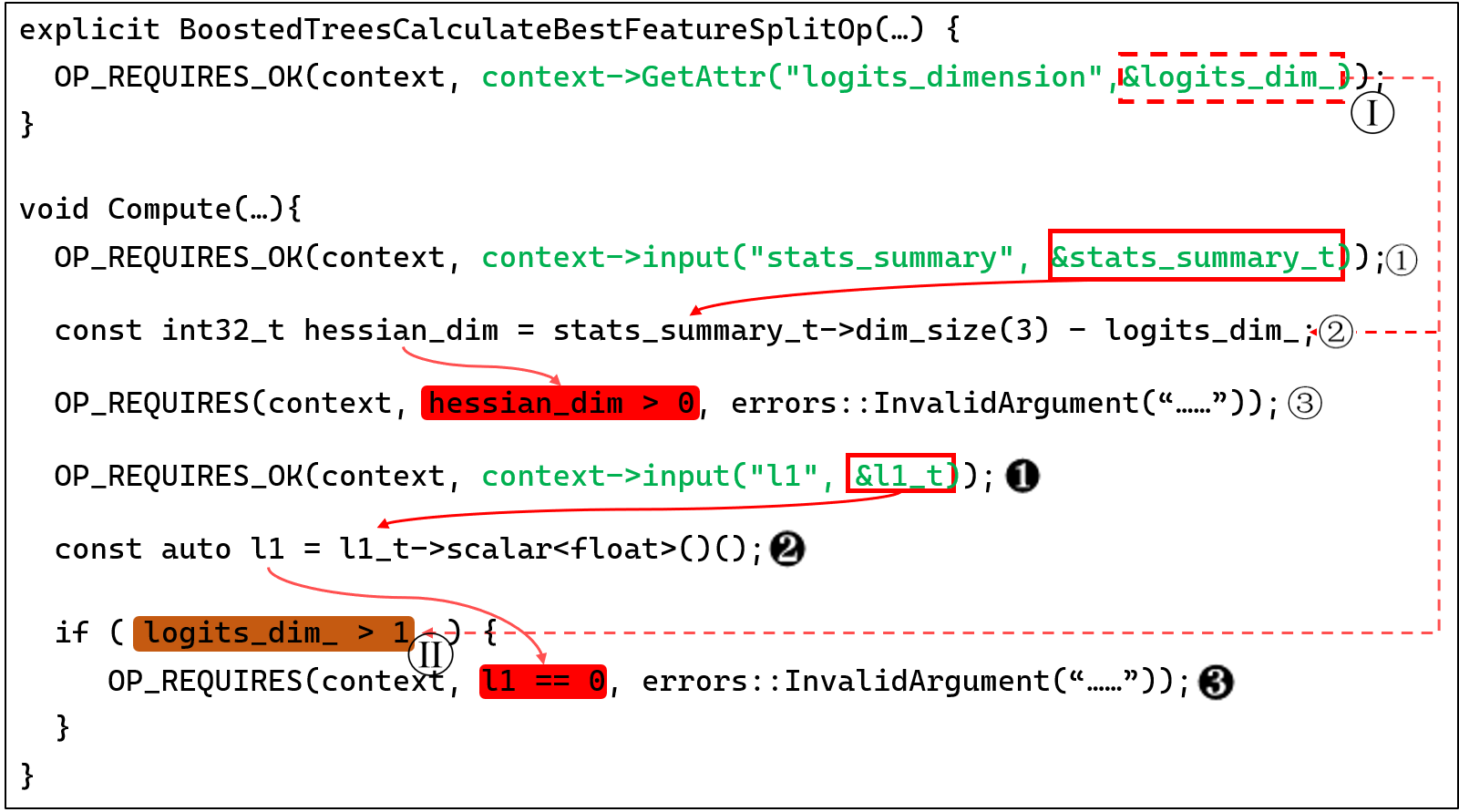} 
\captionsetup{font={small}}
\caption{BTCBFS Constraints Extraction Example} 
\label{BTCBFS Taint Code} 
\vspace{-0.4cm}
\end{figure}

Identifying taint sinks is a critical aspect of the taint analysis method used in this approach. After extensive analysis, it was found that most machine learning frameworks utilize macros to evaluate the validity of operator parameters within the source code. Examples include \texttt{OP\_REQUIRES} in TensorFlow, \texttt{TORCH\_CHECK} in PyTorch, and \texttt{PADDLE\_ENFORCE\_EQ} in Paddle. The operator BTCBFS employs the \texttt{OP\_REQUIRES} macro to determine the relationship between the \texttt{stats\_summary} and \texttt{logits\_dimension} parameters. If the test data fails to meet the constraints, an error is reported and the process is terminated.


The macro's second parameter is an expression, as indicated by the red background in Figure \ref{BTCBFS Taint Code}, while the third parameter is an error output statement. When the expression is false, an output function is called to print the error statement. In reality, the evaluation of an expression in IR is represented as a conditional jump instruction, with its jump target basic block containing an error output function or a check function. As a result, taint sinks are identified based on the following three characteristics:
 
\begin{enumerate}[itemsep=1.3pt,topsep=-0pt,parsep=-0pt]
\item It is a conditional jump instruction.
\item The jump condition contains tainted variables.
\item There is either an error output function or a check function in the jump target basic block.
\end{enumerate}

At taint sinks that satisfy the above characteristics, the jump conditions are extracted as operator constraints. Finally, simplify and revise the extracted constraints to a readable form.

\begin{algorithm}[!ht]\small
    \caption{Constraints Extraction}
    \label{ice}
    \LinesNumbered
    \KwIn {taint and statement}
    \KwOut {constraints of parameters: C}

    \SetKwFunction{FHandleIf}{HandleIf}
    \SetKwFunction{FHandleFor}{HandleFor}
    \SetKwProg{Fn}{Function}{:}{}
  
    \Fn{\FHandleFor{$taint$, $stmt$}}{ 
      \tcp{If taint in ForStmt's body, and the condition in taints, add the condition to tree}
        \If{taint in stmt.body.sink}{
            C.add(convertToCons(stmt.body.sink))
        }
    }
  
    \Fn{\FHandleIf{$taint$, $stmt$}}{ 
        \tcp{If taint in IfStmt's body, add the condition and the sink to tree}
        \If{stmt.body.sink}{
            C.add(convertToCons(stmt.body.sink))
            
            C.add(convertToCons(stmt.cond))
        }
    }

\end{algorithm}



Validation constraints, as indicated by the red background in Figure \ref{BTCBFS Taint Code}, are related to parameter' validation checking. As demonstrated in the example in Figure \ref{BTCBFS Taint Code}, \ding{174} represents validity detection. If the detection fails, the subsequent calculation functions cannot proceed as expected.

For validation constraints, ConFL not only extracts the topmost linear sequence but also analyzes the loop structure, as demonstrated in Algorithm \ref{ice}. When it is determined that the loop body contains only valid detection statements, these validity statements are extracted as constraints.

\subsubsection{Logical Constraints}
We refer to the constraints derived from an if-else branch statement in operators as logical constraints. Logical constraints(brown background in Figure \ref{BTCBFS Taint Code}) are more present in branch judgment. For example, The detection at \textcircled{\uppercase\expandafter{\romannumeral2}} is a logical judgment and is located within the branch judgment, which is related to the specific code logic function.





ConFL adds support for logical constraints by constructing a constraint tree. As in the case of \textcircled{\uppercase\expandafter{\romannumeral2}} in the example, ConFL first determines whether there is a taint in the if statement. If so, it adds the constraint to the constraint tree and then analyzes the legal judgment statement within the if statement block.

Using a constraint tree, ConFL can choose one of the branches to generate fuzzing templates. However, the extracted constraints are at the IR level, which corresponds to the backend C/C++ code. Since ConFL directly calls the Python frontend interface, Python-level constraints are needed as guidance for data generation. In other words, there is a gap between IR constraints and Python parameters. Therefore, it is crucial to elevate the IR constraints to the Python frontend form, making them easily recognizable during fuzzing.

Consider the LARM operator as an example; constraints generated by ConFL are illustrated in Figure \ref{cons_info_LARM}.

\begin{code}
\setlength{\abovecaptionskip}{0cm}
\setlength{\belowcaptionskip}{0.15cm}
\begin{lstlisting}[language=Python]
len(ckpt_path_t) == 1
len(row_remapping.shape()) == 1 
len(row_remapping) == num_rows 
len(col_remapping) == num_cols

\end{lstlisting}
\captionsetup{font={footnotesize}}
\captionof{figure}{Constraint information of the operator LARM.}\label{cons_info_LARM}
\vspace{0.2cm}
\end{code}


The constraints above include both shape requirements of parameters and dependencies between parameters. For instance, \texttt{row\_remapping} must be one-dimensional, and its length should equal the value of \texttt{num\_rows}.

\subsection{Fuzzing Template Generation}


Based on operator information and operator constraints, ConFL generates operator fuzzing templates. These templates do not contain specific fuzzing data for the operator parameters but instead build a test skeleton. In the actual test process, ConFL selects corresponding test data according to the template. The templates are divided into control templates and data templates based on their functions.


\textbf{Control Template.} The control template primarily sets the operator's executable environment, parameter position, parameter type, and other information. By performing topological sorting according to the constraint tree, ConFL first generates a single-parameter template and then creates other parameter templates that depend on this parameter.


Furthermore, when generating control templates, we propose a grouping test for data multiplexing. This is because different Python interfaces may share the same C/C++ backend in TensorFlow. For example, both \texttt{ArgMax} and \texttt{ArgMin} in Python correspond to \texttt{ArgOp} in C++. This is due to the registration mechanism in ML frameworks, which adds various operators, such as \texttt{REGISTER\_OPERATOR} for PaddlePaddle, \texttt{REGISTER\_PRIMITIVE\_EVAL\_IMPL} for MindSpore, and \texttt{REGISTER\_KERNEL\_BUILDER} for TensorFlow. With such a registration method, ConFL establishes correspondence between operators in different languages, enabling parameter data reuse. As previously mentioned, \texttt{ArgMax} and \texttt{ArgMin} share the same parameters: \texttt{input}, \texttt{dimension}, and \texttt{output\_type}.

\textbf{Data Template.} First, ConFL generates a data template and fills it with parameter information in the form of name-value pairs.  ConFL replaces the placeholder with a symbol of the corresponding shape or type according to the explicit information extracted and generates specific values based on the symbol.


By saving the shape and type symbols representing the data, we categorize the generated data to prevent creating too many duplicate parameters. Given the numerous computational steps in ML frameworks, ConFL selects values from a special value set (e.g., boundary value, zero, big integer) when generating specific values. This approach reduces the range of generated parameters and prevents different data from executing the same path while preserving vulnerability detection capability. ConFL then verifies if the parameters satisfy the constraints. If not, it takes targeted modification measures, making simple modifications to the shape or value while retaining the original data characteristics. This lightweight approach saves effort compared to regenerating. Since parameters are checked and modified by explicit and implicit constraints, the operator execution success rate significantly improves.


Based on the constraints in Figure \ref{cons_info_LARM}, ConFL generates a template containing "'col\_remapping': [DI]*num\_rows". "DI" is a data template conforming to the parameter col\_remapping type, representing the use of integer numbers in specific tests. The length of this parameter must equal num\_rows. By applying this template, the following test data in Figure \ref{Test_data_for_LARM} can be generated.

\begin{code}
\setlength{\abovecaptionskip}{0cm}
\setlength{\belowcaptionskip}{0.15cm}
\begin{lstlisting}[language=Python]
  para = { 
    'ckpt_path': 'bundle_checkpoint',
    'old_tensor_name': 'some_scope/matrix',
    'row_remapping': [1],
    'col_remapping': [2147483649]*1073741824,
    'initializing_values': [],
    'num_rows': 1,
    'num_cols': 1073741824,
    'max_rows_in_memory': -1,
  }
\end{lstlisting}
\captionsetup{font={footnotesize}}
\captionof{figure}{Parameters generated for the operator LARM based on the constraints.}\label{Test_data_for_LARM}
\end{code}

\section{Evaluation}
\subsection{Implementation}

The operator collection is implemented using 1K lines of Python code, which parses the operator description and analyzes the source code.  In the process of obtaining interfaces, we modified the Python interpreter - CPython, to monitor the function call chain and determine whether the interface calls C functions. We chose to modify CPython to determine if a C function is called rather than analyzing pybind11 because some interfaces use SWIG, and different function names can be passed to the same C interface, such as the function \texttt{TFE\_Py\_FastPathExecute}.

In the constraint extraction part, we use 500 lines of Python code to extract environmental constraints and dependency constraints. To extract validation constraints and logical constraints, we use 1K lines of C++ code to implement path-insensitive taint analysis based on LLVM. 

Additionally, we use 2K lines of Python code to implement operator test template generation and operator test input generation.

This section evaluates TensorFlow 2.8 using the method introduced in Chapter 3, primarily focusing on the following four aspects:

\begin{enumerate}[leftmargin=*,label=\textbf{RQ\arabic*.}]
\item How effective is ConFL in collecting operators?

\item Are operator constraints helpful for parameter generation?

\item Can ConFL find vulnerabilities in real-world applications?
\end{enumerate}

The machine used for running the experiments is equipped with Intel Xeon E5-2630 2.20 GHz CPU, Tesla P4 GPU, 128GB RAM, Ubuntu 20.04 LTS, and Python3.8.

\subsection{Effectiveness of Operator Collection}

Unlike collecting operator information from documents, ConFL collects operators by analyzing the codes of ML framework itself, and the operator can be directly called for fuzzing. When adapting to the newest version, ConFL automatically extracts operators of the version without re-collect public code segments or re-analyze operator documents.

ConFL primarily tests the raw\_ops module, consisting of 1,355 operators. Out of these, 24 operators are deprecated or meaningless, like the \texttt{Abort} operator, leaving 1,331 valid operators in the raw\_ops module for testing.

Out of the remaining 1,331 operators, 65 depend on TPU, including operators like \texttt{SendTPUEmbeddingGradients}. Although ConFL is theoretically capable of detecting these operators, hardware limitations prevented their inclusion in our experiment. Consequently, we  selected 1,266 non-TPU-dependent operators as test targets.

Additionally, other modules can be tested by ConFL, such as the IO module, where CVE-2020-26269 was found. 

\begin{figure}[htbp]
\setlength{\abovecaptionskip}{0.15cm}
\centering
\includegraphics[width=0.35\textwidth]{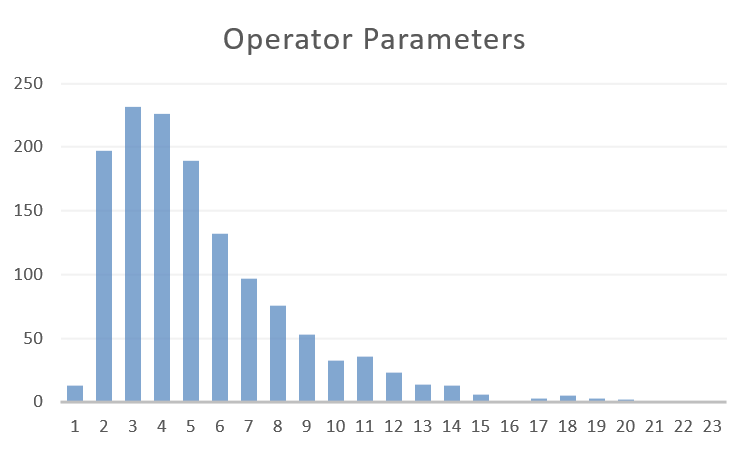} 
\captionsetup{font={footnotesize}}
\caption{Statistics of operator parameters} 
\label{Soop}
\vspace{-0.5cm}
\end{figure}

Figure \ref{Soop} displays the distribution of operator parameter counts, with 976 operators having 2 to 6 parameters. The most common count is 3 parameters, found in 232 operators. Four operators have over 20 parameters, and on average, each operator has 5 parameters.

\vspace{0.4cm}
\textbf{Answer to RQ1:} ConFL collects operators with a total number of 1,355, of which 1,331 operators are valid. After excluding operators that rely on hardware, We randomly select 400 of the 1,266 operators as test targets. 

\subsection{Effectiveness of Operator Constraints Extraction}

\begin{table}[!ht]
    \setlength{\abovecaptionskip}{0.15cm}
    \setlength{\belowcaptionskip}{-0.cm}
    \centering
    \captionsetup{font={small}}
    \caption{Constraints Counts} \label{compare}
    \begin{tabular}{cccccc}
        \Xhline{1pt}
             Constraints &  Counts \\ 
        \midrule
            Environmental Constraints & 6  \\ 
            Dependency Constraints & 23   \\ 
            Validation Constraints & 1,519   \\ 
            Logical Constraints & 98  \\ 
            
        \Xhline{1pt}
    \end{tabular}
    \vspace{-0.5cm}
\end{table}

We collect 6 environmental constraints through expert experience. Although the number of environmental constraints is relatively small, they are very effective. When compared to tests that do not apply these constraints, more new code can be executed, and vulnerabilities can be found. ConFL extracts 23 dependency constraints after analyzing the operator's information. By using dependency constraints, some operators can be executed successfully. The success of an operator's execution is directly related to whether the parameters can pass validation verification. Due to the complexity of the types and numbers of operators, 1519 validation constraints are extracted. Similarly, the introduction of functional constraints allows us to build a complete constraint tree that covers a sufficient amount of code.

\begin{table}[!ht]
    \setlength{\abovecaptionskip}{0.15cm}
    \setlength{\belowcaptionskip}{-0.cm}
    \centering
    \captionsetup{font={small}}
    \caption{Validation Constraints Classification of a Single Parameter} \label{single-cons}
    \begin{tabular}{cccccc}
        \Xhline{1pt}
             ndim &  shape & size & value & dtype\\ 
        \midrule
            837 & 92 & 202 & 92 & 15  \\ 
        \Xhline{1pt}
    \end{tabular}
    \vspace{-0.5cm}
\end{table}

\begin{table}[!ht]
    \setlength{\abovecaptionskip}{0.15cm}
    \setlength{\belowcaptionskip}{0.2cm}
    \centering
    \captionsetup{font={small}}
    \caption{Validation Constraints Classification among Parameters } \label{multi-cons}
    \begin{tabular}{cccccc}
        \Xhline{1pt}
             & ndim &  shape & size & value \\ 
        \midrule
            ndim & 1  \\ 
            shape & 10 & 199  \\ 
            size & 3 & 16 & 12 \\ 
            value & 11 & 17 & - & 12  \\ 
        \Xhline{1pt}
    \end{tabular}
    \vspace{0.1cm}
\end{table}
We classify the validation constraints into two types: constraints related to a single parameter and constraints among parameters.  Moreover, we describe a parameter from various perspectives. \texttt{ndim} represents  the number of dimensions, \texttt{shape} refers to each dimension of a tensor, \texttt{size} is the number of elements, \texttt{value} describes the specific value, while \texttt{dtype} indicates the type of a parameter. As shown in Table \ref{single-cons}, the ndim-type has the largest proportion of single parameter constraints. For example, in \texttt{ArgMax}, we obtain the constraint ``dimension.ndim == 0" to restrict the ndim of parameter \texttt{dimension}. In Table \ref{multi-cons}, the rows and columns specify the attribute constraints among parameters. For instance, the ``10" indicates that there are 10 constraints between shape and ndim, such as ``input.ndim > block\_shape.shape[0]" for operator \texttt{BatchToSpaceND}.

\subsection{Effectiveness of Constraint-guided operator input Generation}
In this experiment, we set up two comparison on code coverage: Compared with random generation (Atheris) and compared with state-of-the-art(SOTA) fuzzers. With the consideration of various SOTA fuzzers can not cover all opertors, we first conduct a experiment on comparison with Atheris in 1,266 operators. Then we select 400 operators that all the SOTA fuzzers can cover commonly, then conduct another experiment on comparison with SOTA fuzzers.

\noindent\textbf{Compared with random generation.} We separately employ Atheris and ConFL to generate 10,000 test inputs for each operator, and record the number of successful executions. We define a successful execution as one that triggers either a crash or normal exit, while an unsuccessful execution is one that fails due to a parameter error, such as a Python code exception. The test result show that Atheris achieves a total of 669,249 successful execution times for all tested operators, while ConFL reaches 3,534,170 times. The increase rate amounts to 428.08\%, demonstrating that ConFL significantly improves the validity of the generated inputs.

\begin{figure}[htbp]
\setlength{\abovecaptionskip}{0.15cm}
\centering
\includegraphics[width=0.35\textwidth]{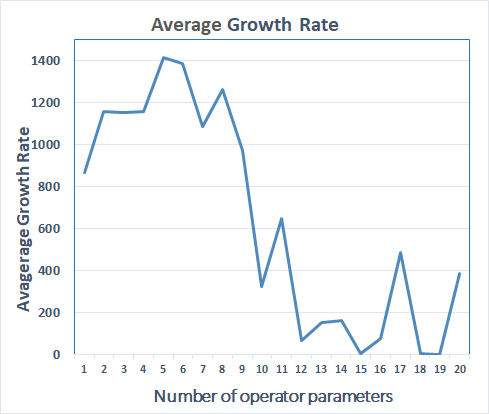} 
\captionsetup{font={footnotesize}}
\caption{Average Growth Rate} 
\label{OPSb}
\vspace{-0.5cm}
\end{figure}

Furthermore, we examined the relationship between the increase rate and the number of parameters. We organized the operators based on their parameter count and assigned an ID to each. Figure \ref{OPSb} illustrates the increase rate of ConFL compared to Atheris. The average increase rate for all operators is 625.94\%; operators with 5 parameters experience the highest growth rate at 1,417.94\%. Although the increase rate declines as the number of parameters grows, ConFL still outperforms Atheris significantly. This suggests that when an operator has few parameters, it can be successfully executed using random data generation. However, as the number of parameters rises, the limitations of random generation become more evident, and the benefits of constraint-based generation grow more pronounced.

\begin{figure}[htbp]
\setlength{\abovecaptionskip}{0.15cm}
\centering
\includegraphics[width=0.35\textwidth]{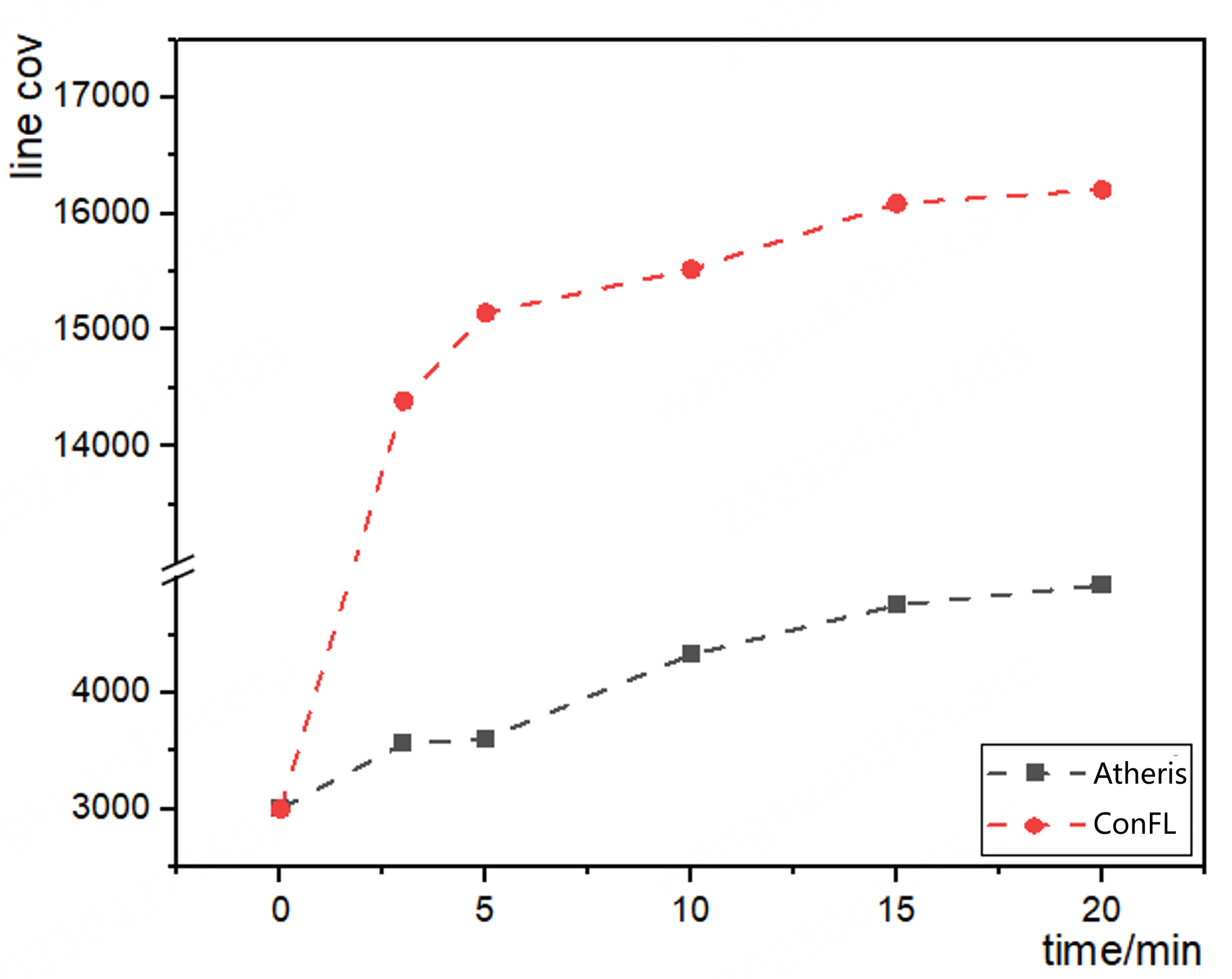} 
\captionsetup{font={footnotesize}}
\caption{Coverage with constraint} 
\label{Operator_coverage1}
\vspace{-0.4cm}
\end{figure}

Concerning code coverage, Figure \ref{Operator_coverage1} displays the results. We fuzzed 1,266 operators for 20 minutes. Through the code coverage analysis, we discovered that the coverage state stabilizes at 20 minutes, with Atheris covering only 4,929 lines of code. This suggests that the majority of inputs are invalid, causing stagnation in the validation checking. In contrast, using constraints, ConFL's code coverage not only increased rapidly in the first 5 minutes but also sustained steady growth during the subsequent testing. Within the limited time, ConFL increased the coverage by 228.83\% compared to Atheris, demonstrating the efficiency of ConFL in generating valid inputs.

\begin{figure}[htbp]
\setlength{\abovecaptionskip}{0.15cm}
\centering
\includegraphics[width=0.45\textwidth]{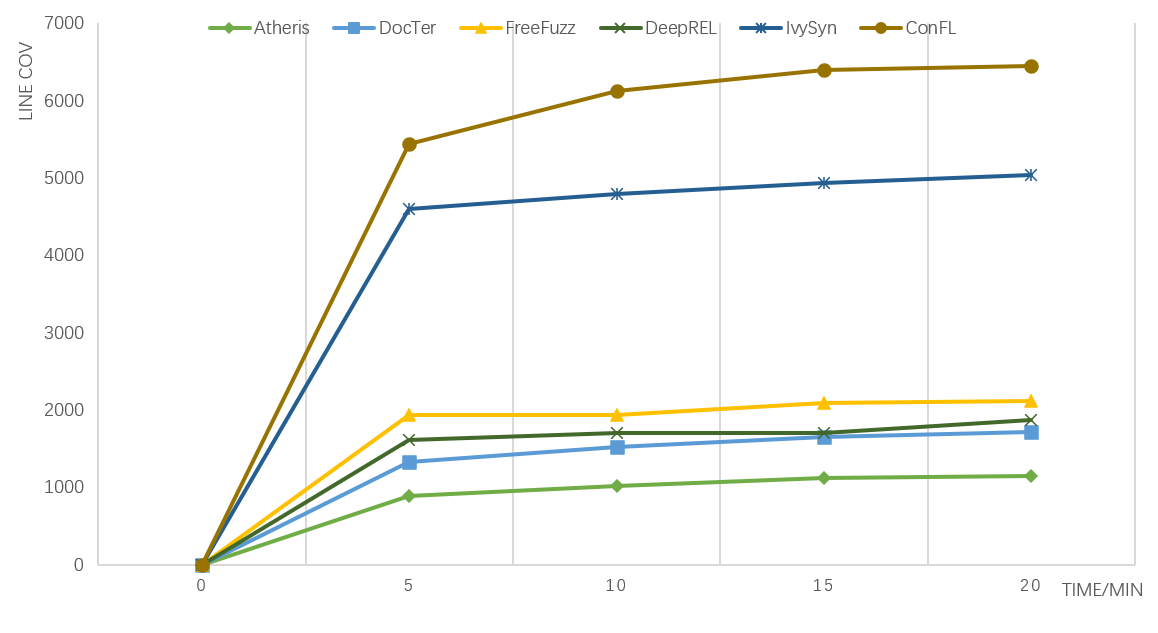} 
\captionsetup{font={footnotesize}}
\caption{Comparison on Code Coverage with SOTA Fuzzers} 
\label{Operator_coverage}
\vspace{-0.4cm}
\end{figure}

\noindent\textbf{Compared with state-of-the-art fuzzers.} We compare ConFL with state-of-the-art (SOTA) fuzzers, including DocTer, FreeFuzz, DeepRel, and IvySyn. Since some SOTA fuzzers cannot cover all 1,226 operators, we select 400 operators that all the SOTA fuzzers can commonly cover for fairness as the benchmark, using Atheris as the baseline. Regarding test time settings, we utilize all the fuzzers to test each operator in the benchmark for 20 minutes. Subsequently, we record the total code coverage of all operators in the benchmark.

As depicted in the figure\ref{Operator_coverage}, the results of the testing show that ConFL consistently outperforms DocTer, FreeFuzz, DeepRel, and IvySyn in code coverage metrics. The figure illustrates the code coverage achieved by each fuzzer, with ConFL achieving significantly higher coverage compared to its counterparts. This indicates that ConFL is more effective in generating valid inputs and exploring a broader range of code paths. 

\vspace{0.4cm}
\textbf{Answer to RQ2:} Constraints are helpful for generating valid inputs of operators, and largely improve the success rate of execution. Additionally, our constraint-based approach significantly increases code coverage of operators compared with state-of-the-art fuzzers.

\subsection{Effectiveness of Vulnerability Detection}

The vulnerability detection results of ConFL when applied to the TensorFlow framework are presented in Table \ref{tableVI}. ConFL successfully identified a total of 84 vulnerabilities within the TensorFlow framework, all of which have been confirmed and assigned CVE numbers. A selection of representative vulnerabilities is detailed in the table, while a comprehensive list can be found in \cite{b42}.

\begin{table}[!ht]\footnotesize
    \centering
        \setlength{\abovecaptionskip}{0.15cm}
    \setlength{\belowcaptionskip}{-0.1cm}
    \captionsetup{font={small}}
    \caption{TOP 5 vulnerability type of TensorFlow Framework }
    \label{tableVI}
    \begin{tabular}{cccc}
    \Xhline{1pt}
\textbf{Framweork} & \textbf{Type} & \textbf{Example Operator} & \textbf{Example CVE} \\ \hline
\multirow{5}{*}{Tensorflow}   & OOB & SparseBincount& CVE-2021-41226        \\
    & NPE& Conv2D & CVE-2021-41209        \\
    & FPE& AvgPoolGrad & CVE-2022-21725        \\
    & IOF & StringNGrams& CVE-2022-21733        \\
    & UAF& BoostedTreesCreateEnsemble& CVE-2021-37652        \\ \Xhline{1pt}
\end{tabular}

    \vspace{0.1cm}
\end{table}

\noindent\textbf{Vulnerability Type Analysis.} As depicted in Table \ref{tableVI}, the 84 discovered vulnerabilities are classified according to their types, with the top five types being Out of Bound (OOB), Null Pointer Exception (NPE), Floating Point Exception (FPE), Integer Overflow (IOF), and Use After Free (UAF). Here, we provide a brief overview of each vulnerability type along with corresponding examples:

\begin{itemize}
\item Out of Bound (OOB): OOB vulnerabilities occur when an operation accesses memory outside of its intended bounds, potentially leading to data corruption, crashes, or security breaches. One such example is CVE-2021-41226, which corresponds to the TensorFlow operator \texttt{SparseBincount}.

\item Null Pointer Exception (NPE): NPE vulnerabilities arise when a program attempts to access or manipulate an object via a null pointer reference, potentially causing unexpected behavior, crashes, or security issues. A notable instance is CVE-2021-41209, associated with the TensorFlow operator \texttt{Conv2D}.

\item Floating Point Exception (FPE): FPE vulnerabilities involve errors in floating-point operations, such as division by zero or overflow, which can result in crashes or incorrect calculations, impacting the system's reliability. An example of an FPE vulnerability is CVE-2022-21725, related to the TensorFlow operator \texttt{AvgPoolGrad}.

\item Integer Overflow (IOF): IOF vulnerabilities occur when an integer operation produces a value too large or too small to be represented by the integer type, potentially leading to data corruption, crashes, or other unintended consequences. An instance of an IOF vulnerability is CVE-2022-21733, corresponding to the TensorFlow operator \texttt{StringNGrams}.

\item Use After Free (UAF): UAF vulnerabilities happen when a program continues to use a memory object after it has been freed, potentially resulting in crashes, data corruption, or security exploits. An example of a UAF vulnerability is CVE-2021-37652, linked to the TensorFlow operator \texttt{Boosted\hyp{}TreesCreateEnsemble}.
\end{itemize}

In summary, the application of ConFL to the TensorFlow framework led to the identification of 84 vulnerabilities, spanning a range of types. By understanding and addressing these vulnerabilities, developers can work towards enhancing the security, stability, and reliability of the TensorFlow framework.

\noindent\textbf{Causality Analysis.} By analyzing 84 vulnerabilities detected by ConFL in TensorFlow, we have identified the causality of these vulnerabilities and classified them into three categories: shape, type, and value, as displayed in Table \ref{vul_type}.

Regarding shape, a zero-dimensional vector may lead to NPE, OOB, and FPE vulnerabilities, such as CVE-2021-37672. Alternatively, a large value may cause OOB and NPE vulnerabilities, exemplified by CVE-2021-37655. In terms of type, an incorrect tensor type value can trigger Denial of Service (DoS) vulnerabilities, as seen in CVE-2020-26268. Concerning value, tensor data or parameter values of zero can result in FPE and OOB vulnerabilities, such as CVE-2022-21725; large integer values can lead to OOB, IOF, and Type Confusion (TC) vulnerabilities, as in the case of CVE-2022-21727; and negative values can cause OOB and IOF vulnerabilities, as demonstrated by CVE-2022-21733.

\begin{table}[!ht]\footnotesize
    \setlength{\abovecaptionskip}{0.15cm}
    \setlength{\belowcaptionskip}{-0.cm}
    \centering
    \captionsetup{font={small}}
    \caption{Causality of Vulnerabilities} \label{vul_type}
    \begin{threeparttable}
    \begin{tabularx}{0.45\textwidth}{cllc}
        \Xhline{1pt}
        \textbf{Category} & \textbf{PoC Input} & \textbf{Vulnerability Type} & \textbf{Example CVE} \\ 

        \midrule
        \multirow{2}{*}{Shape}
         & zero dim  & NPE, FPE, OOB & CVE-2021-37672 \\ 
         & big index  & OOB, NPE & CVE-2021-37655 \\ 
        \midrule
        \multirow{1}{*}{Type}
         & string & DoS & CVE-2020-26268 \\
        \midrule
        \multirow{3}{*}{Value}
         & zero & FPE, OOB & CVE-2022-21725 \\ 
         & big int & OOB, IOF, TC & CVE-2022-21727 \\ 
         & negative & OOB, IOF & CVE-2022-21733 \\ 
        \Xhline{1pt}
    \end{tabularx}
    \vspace{0.1cm}
    \end{threeparttable}
\end{table}

We find that current ML frameworks prioritize performance and functionality over security, lacking comprehensive user input validation, particularly for empty arrays and empty handlers. Additionally, the computational nature of machine learning algorithms results in frequent floating-point and integer overflow issues. Lastly, the interdependent relationships between ML framework operator parameters may generate valid parameters that still impact other parameters and cause computational problems.

\noindent\textbf{Case Study.} In the following, we discuss a vulnerability example to illustrate how ConFL can effectively generate valid inputs, enabling efficient detection of vulnerabilities in real-world ML frameworks.

ConFL identified an out-of-bound read vulnerability in the BTCBFS operator. The vulnerability Proof of Concept (PoC) demonstrates that the parameter \texttt{split\_type} is a string with only two valid values: \texttt{inequality} or \texttt{equality}. Simultaneously, there is a validation constraint between the \texttt{stats\_summary} and \texttt{logits\_dimension} parameters: the dimension of \texttt{stats\_summary} is related to the value of \texttt{logits\_dimension}. ConFL continually generates valid parameters based on operator constraints to probe deeper vulnerabilities in the code. In this example, considering shape, type, and value constraints, the parameter range of \texttt{split\_type} is limited. ConFL also generates valid data for \texttt{stats\_summary} and \texttt{logits\_dimension} parameters, utilizing the constraints. These methods help avoid wasting time and computational resources on shallow code.

\begin{code}
\setlength{\abovecaptionskip}{0cm}
\setlength{\belowcaptionskip}{0.15cm}
\begin{lstlisting}[language=Python]
tensorflow.raw_ops.BoostedTreesCalculateBestFeatureSplit(
    node_id_range=[0x400000,0x400001],
    stats_summary=[
        [[[2.0, 3.0]], [[3., 3.]], [[3., 3.]]], 
        [[[3., 4.]], [[5., 6.]], [[6., 6.]]]
    ],
    l1=[0.0],
    l2=[0.0],
    tree_complexity=[1.0],
    min_node_weight=[0.7],
    logits_dimension = 1,
    split_type = 'equality'
)
\end{lstlisting}
\captionsetup{font={footnotesize}}
\captionof{figure}{PoC for BTCBFS.}
\end{code}

Finally, ConFL dicovers this vulnerability located in BTCBFS operator with a boundary value of the parameters, which leads to an out-of-bound access. As shown in the source code below, the parameter \texttt{node\_id} takes the value of the input parameter \texttt{node\_id\_range} , which may exceed the range of \texttt{stats\_summary}. Then, the pointer of \texttt{stats\_mat} will point to an out-of-control address.

\begin{code}
\setlength{\abovecaptionskip}{0cm}
\setlength{\belowcaptionskip}{0.15cm}
\begin{lstlisting}[language=C]
ConstMatrixMap stats_mat(&stats_summary(node_id, 0, 0, 0), ...);
  
const Eigen::VectorXf total_grad =
  stats_mat.leftCols(logits_dim).colwise().sum();
\end{lstlisting}
\captionsetup{font={footnotesize}}
\captionof{figure}{Source code that cause the vulnerability in BTCBFS.}
\end{code}

\noindent\textbf{Vulnerabilites in Other ML Frameworks.} Despite being in its early prototype stage, we have attempted to extend ConFL to test other ML frameworks, including PyTorch and PaddlePaddle. To date, ConFL has discovered 7 vulnerabilities across these platforms. In PaddlePaddle, ConFL identified a total of 4 vulnerabilities, while in PyTorch, it detected 3 out-of-bound (OOB) vulnerabilities. These results demonstrate that ConFL exhibits strong adaptability to other ML frameworks.

\begin{table}[!ht]\footnotesize
    \setlength{\abovecaptionskip}{0.1cm}
    \setlength{\belowcaptionskip}{-0.cm}
    \centering
    \captionsetup{font={small}}
    \caption{Detected Vulnerabilities in other ML Frameworks}
    \label{tableVI1}
    \begin{tabular}{cccc}
    \Xhline{1pt}
\textbf{Framework} & \textbf{Type} & \textbf{Operator} & \textbf{ISSUE-ID} \\ \hline
\multirow{4}{*}{PaddlePaddle}& \multirow{2}{*}{OOB}  & gather\_tree  & 33382      \\
  && strided\_slice & 33006                 \\
  & FPE& Pool3d & 33036                 \\
  & DF& split  & 32942    \\ \hline
\multirow{3}{*}{Pytorch}  & \multirow{3}{*}{OOB} & quantized\_lstm\_cell & 50037                 \\
     &  & \_remove\_batch\_dim   & 50038                 \\
    & & native\_layer\_norm  & 50090     \\ \Xhline{1pt}
\end{tabular}

    \vspace{0.1cm}

\end{table}

\vspace{0.4cm}
\textbf{Answer to RQ3:} ConFL can extract the constraints between multiple parameters, and generate valid parameters, which is effective for discovering vulnerabilities of ML frameworks in the real world.

\section{Discussion}
In this section, we present the limitations and possible solutions of improvement in future.

\noindent\textbf{Adaptability to other ML frameworks.} 
The design of ConFL can be effortlessly adapted to other ML frameworks with minimal modifications. For constraint extraction, ML frameworks like PyTorch and PaddlePaddle employ macros for parameter validation in the source code, similar to TensorFlow. For instance, PyTorch defines operators in native\_functions.yaml and Declarations.cwrap, and verifies parameter validity in the source code using TORCH\_CHECK. These files can be parsed to extract constraints as well.

Since the reflection mechanism is compatible with all frameworks featuring a Python frontend, ConFL can automatically generate templates for such frameworks. Moreover, ConFL can produce parameters tailored to the specific characteristics of each ML framework, such as custom types. This adaptability allows ConFL to be a versatile solution for various machine learning frameworks.

\noindent\textbf{Optimization of constraint solving.} With the extracted constraints, efficient constraint solving techniques can be employed to generate valid test inputs more effectively. This can potentially lead to a higher coverage of the operator code and an increased likelihood of finding deep bugs.

\noindent\textbf{Integration with other fuzzing techniques.} The constraint extraction technique can be combined with other fuzzing techniques, such as grammar-based fuzzing or coverage-guided fuzzing, to achieve a more comprehensive and effective fuzz testing process.

\noindent\textbf{Operator Optimization.} ML frameworks also focus on the optimization such as operator fusion~\cite{b36} in the practical computation process, which aims to reduce the occupation of memory and improve the efficiency. At present, ConFL tests operators separately and the fused operator should be considered in the future.

\noindent\textbf{File mutation.} Some operators require specific file formats as parameters, such as an image file for \texttt{DecodeJEPG} or an audio file for \texttt{DecodeWAV}. To support complex operators with composite types, we plan to add file format mutation in future work.
\section{Related Work}

\subsection{Fuzzing System and Application Interfaces}
Some previous work focusing on fuzzing various system and interfaces, including cloud service APIs \cite{b44} , OS kernel interfaces \cite{b46,b47,b48}, and native library interfaces \cite{b49}. For example, NTFuzz \cite{b22} is a type-aware Windows kernel fuzzing framework,  which can automatically infer system call types on Windows on a large scale. APICRAFT \cite{b23} utilizes static and dynamic information to gather control and data dependencies of API functions. And it employs a multi-objective genetic algorithm to combine the collected dependencies and build a high-quality fuzzy driver. 

Although there are similarities between system interfaces and machine learning APIs, previous fuzzing tools are not directly applicable to fuzzing machine learning APIs for two main reasons. First, machine learning APIs utilize domain-specific data types, such as tensors, which necessitate specialized fuzzing techniques. Second, machine learning APIs exhibit unique constraints and interdependencies between parameters, which general system interface fuzzing tools may not effectively handle.

\subsection{Fuzzing ML Framework}
In recent years, researchers have made major strides in the fuzzing ML frameworks. Q. Xiao et al. \cite{b9}, X. Tan et al. \cite{b10} studied the security issues in the third-party dependency libraries of ML frameworks, but did not pay more attention to the security of the source code of ML frameworks in depth.

Xie et al. \cite{b14} proposed DeepHunter, a general-purpose fuzz testing tool for deep learning frameworks, using scalable coverage criteria and a seed selection strategy. However, their random mutation at the operator level lacks constraint or verification, reducing the legitimacy of samples and automation efficiency. Luo et al. \cite{b15} proposed operator-level automated testing for deep learning frameworks using the Monte Carlo tree search algorithm and combining model-level and source-level mutation. Wang et al.~\cite{b16}  studied the effectiveness of unit test generation techniques for machine learning libraries, finding that most existing libraries lack high-quality unit test suites. The uncovered code is primarily due to insufficient valid parameters for tests, leading them to propose a future direction combining test generation and parameter analysis.

DocTer \cite{b11} analyzes API documentation to extract input constraints for machine learning API functions. While this approach can provide constraints for some functions, its effectiveness is limited by the completeness and accuracy of the documentation. 

FreeFuzz \cite{b12} fuzzes DL libraries by mining open-source code/models, automatically running them with instrumentation, and using the traced dynamic information for fuzz testing. However, it lacks systematic testing procedures for operators.

DeepRel~\cite{b40} extends FreeFuzz by leveraging function similarity to transfer inputs between test cases. It uses function signatures and documentation to generate valid inputs for some functions, but may be limited when documentation is lacking.

IvySyn~\cite{b41} is a specialized tool for detecting vulnerabilities in DL kernel code. It identifies DL kernel implementations and performs mutation-based fuzzing with type-aware mutations. IvySyn uses developer test suites as initial test cases, sharing similar limitations with FreeFuzz and DeepRel.

Our approach not only focuses on achieving higher code coverage but also ensures that the generated test inputs are valid and conform to the constraints of the target ML frameworks. By automatically extracting input constraints from the source code of operators, ConFL can generate a more comprehensive and accurate set of test inputs. This, in turn, improves the efficiency and effectiveness of the fuzzing process in identifying vulnerabilities.

\section{Conclusion}

In this paper, we introduce ConFL, an innovative tool designed to generate valid operator parameters for uncovering hidden security vulnerabilities in ML frameworks. Initially, ConFL analyzes the source code to collect operators. It then extracts constraints from the operators' source codes. Finally, ConFL automatically constructs operator test templates and generates test inputs guided by the extracted constraints.Through our evaluation, ConFL demonstrates remarkable proficiency in generating valid parameters. Furthermore, our approach has successfully identified 84 vulnerabilities in TensorFlow and 7 in PyTorch and PaddlePaddle.

\begin{acks}
This work is supported by the National Key R\&D Program of China with No.2020AAA0104300.
\end{acks}

\bibliographystyle{ACM-Reference-Format}
\bibliography{reference}


\begin{thebibliography}{28}


\ifx \showCODEN    \undefined \def \showCODEN     #1{\unskip}     \fi
\ifx \showDOI      \undefined \def \showDOI       #1{#1}\fi
\ifx \showISBNx    \undefined \def \showISBNx     #1{\unskip}     \fi
\ifx \showISBNxiii \undefined \def \showISBNxiii  #1{\unskip}     \fi
\ifx \showISSN     \undefined \def \showISSN      #1{\unskip}     \fi
\ifx \showLCCN     \undefined \def \showLCCN      #1{\unskip}     \fi
\ifx \shownote     \undefined \def \shownote      #1{#1}          \fi
\ifx \showarticletitle \undefined \def \showarticletitle #1{#1}   \fi
\ifx \showURL      \undefined \def \showURL       {\relax}        \fi
\providecommand\bibfield[2]{#2}
\providecommand\bibinfo[2]{#2}
\providecommand\natexlab[1]{#1}
\providecommand\showeprint[2][]{arXiv:#2}

\bibitem[Acharya et~al\mbox{.}(2020)]%
        {b36}
\bibfield{author}{\bibinfo{person}{Aravind Acharya}, \bibinfo{person}{Uday
  Bondhugula}, {and} \bibinfo{person}{Albert Cohen}.}
  \bibinfo{year}{2020}\natexlab{}.
\newblock \showarticletitle{Effective Loop Fusion in Polyhedral Compilation
  Using Fusion Conflict Graphs}.
\newblock \bibinfo{journal}{\emph{{ACM} Trans. Archit. Code Optim.}}
  \bibinfo{volume}{17}, \bibinfo{number}{4} (\bibinfo{year}{2020}),
  \bibinfo{pages}{26:1--26:26}.
\newblock


\bibitem[AIVul(2022)]%
        {b42}
\bibfield{author}{\bibinfo{person}{AIVul}.} \bibinfo{year}{2022}\natexlab{}.
\newblock \bibinfo{title}{vulsinfo}.
\newblock \bibinfo{howpublished}{\url{https://sites.google.com/view/aivul/}}.
\newblock


\bibitem[Atlidakis et~al\mbox{.}(2019)]%
        {b44}
\bibfield{author}{\bibinfo{person}{Vaggelis Atlidakis},
  \bibinfo{person}{Patrice Godefroid}, {and} \bibinfo{person}{Marina
  Polishchuk}.} \bibinfo{year}{2019}\natexlab{}.
\newblock \showarticletitle{Restler: Stateful rest api fuzzing}. In
  \bibinfo{booktitle}{\emph{2019 IEEE/ACM 41st International Conference on
  Software Engineering (ICSE)}}. IEEE, \bibinfo{pages}{748--758}.
\newblock


\bibitem[Babi{\'c} et~al\mbox{.}(2019)]%
        {b49}
\bibfield{author}{\bibinfo{person}{Domagoj Babi{\'c}}, \bibinfo{person}{Stefan
  Bucur}, \bibinfo{person}{Yaohui Chen}, \bibinfo{person}{Franjo
  Ivan{\v{c}}i{\'c}}, \bibinfo{person}{Tim King}, \bibinfo{person}{Markus
  Kusano}, \bibinfo{person}{Caroline Lemieux}, \bibinfo{person}{L{\'a}szl{\'o}
  Szekeres}, {and} \bibinfo{person}{Wei Wang}.}
  \bibinfo{year}{2019}\natexlab{}.
\newblock \showarticletitle{Fudge: fuzz driver generation at scale}. In
  \bibinfo{booktitle}{\emph{Proceedings of the 2019 27th ACM Joint Meeting on
  European Software Engineering Conference and Symposium on the Foundations of
  Software Engineering}}. \bibinfo{pages}{975--985}.
\newblock


\bibitem[Choi et~al\mbox{.}(2021a)]%
        {b46}
\bibfield{author}{\bibinfo{person}{Jaeseung Choi}, \bibinfo{person}{Kangsu
  Kim}, \bibinfo{person}{Daejin Lee}, {and} \bibinfo{person}{Sang~Kil Cha}.}
  \bibinfo{year}{2021}\natexlab{a}.
\newblock \showarticletitle{NTFuzz: Enabling type-aware kernel fuzzing on
  windows with static binary analysis}. In \bibinfo{booktitle}{\emph{2021 IEEE
  Symposium on Security and Privacy (SP)}}. IEEE, \bibinfo{pages}{677--693}.
\newblock


\bibitem[Choi et~al\mbox{.}(2021b)]%
        {b22}
\bibfield{author}{\bibinfo{person}{Jaeseung Choi}, \bibinfo{person}{Kangsu
  Kim}, \bibinfo{person}{Daejin Lee}, {and} \bibinfo{person}{Sang~Kil Cha}.}
  \bibinfo{year}{2021}\natexlab{b}.
\newblock \showarticletitle{NtFuzz: Enabling Type-Aware Kernel Fuzzing on
  Windows with Static Binary Analysis}. In \bibinfo{booktitle}{\emph{2021 IEEE
  Symposium on Security and Privacy (SP)}}. \bibinfo{pages}{677--693}.
\newblock
\urldef\tempurl%
\url{https://doi.org/10.1109/SP40001.2021.00114}
\showDOI{\tempurl}


\bibitem[Christou et~al\mbox{.}(2023)]%
        {b41}
\bibfield{author}{\bibinfo{person}{Neophytos Christou}, \bibinfo{person}{Di
  Jin}, \bibinfo{person}{Vaggelis Atlidakis}, \bibinfo{person}{Baishakhi Ray},
  {and} \bibinfo{person}{Vasileios~P. Kemerlis}.}
  \bibinfo{year}{2023}\natexlab{}.
\newblock \showarticletitle{IvySyn: Automated Vulnerability Discovery in Deep
  Learning Frameworks}. In \bibinfo{booktitle}{\emph{USENIX Security Symposium
  (SEC)}}.
\newblock


\bibitem[Corina et~al\mbox{.}(2017)]%
        {b47}
\bibfield{author}{\bibinfo{person}{Jake Corina}, \bibinfo{person}{Aravind
  Machiry}, \bibinfo{person}{Christopher Salls}, \bibinfo{person}{Yan
  Shoshitaishvili}, \bibinfo{person}{Shuang Hao}, \bibinfo{person}{Christopher
  Kruegel}, {and} \bibinfo{person}{Giovanni Vigna}.}
  \bibinfo{year}{2017}\natexlab{}.
\newblock \showarticletitle{Difuze: Interface aware fuzzing for kernel
  drivers}. In \bibinfo{booktitle}{\emph{Proceedings of the 2017 ACM SIGSAC
  Conference on Computer and Communications Security}}.
  \bibinfo{pages}{2123--2138}.
\newblock


\bibitem[Deng et~al\mbox{.}(2022)]%
        {b40}
\bibfield{author}{\bibinfo{person}{Yinlin Deng}, \bibinfo{person}{Chenyuan
  Yang}, \bibinfo{person}{Anjiang Wei}, {and} \bibinfo{person}{Lingming
  Zhang}.} \bibinfo{year}{2022}\natexlab{}.
\newblock \showarticletitle{Fuzzing deep-learning libraries via automated
  relational API inference}. In \bibinfo{booktitle}{\emph{Proceedings of the
  30th ACM Joint European Software Engineering Conference and Symposium on the
  Foundations of Software Engineering}}. \bibinfo{pages}{44--56}.
\newblock


\bibitem[Google(2021)]%
        {b39}
\bibfield{author}{\bibinfo{person}{Google}.} \bibinfo{year}{2021}\natexlab{}.
\newblock \bibinfo{title}{Tensorflow}.
\newblock
  \bibinfo{howpublished}{\url{https://www.tensorflow.org/api_docs/python/tf/dtypes}}.
\newblock


\bibitem[Google(2022a)]%
        {b20}
\bibfield{author}{\bibinfo{person}{Google}.} \bibinfo{year}{2022}\natexlab{a}.
\newblock \bibinfo{title}{AFL}.
\newblock \bibinfo{howpublished}{\url{https://github.com/google/AFL}}.
\newblock


\bibitem[Google(2022b)]%
        {b27}
\bibfield{author}{\bibinfo{person}{Google}.} \bibinfo{year}{2022}\natexlab{b}.
\newblock \bibinfo{title}{BoostedTreesCalculateBestFeatureSplit}.
\newblock
  \bibinfo{howpublished}{\url{https://www.tensorflow.org/api_docs/python/tf/raw_ops/BoostedTreesCalculateBestFeatureSplit}}.
\newblock


\bibitem[Google(2022c)]%
        {b17}
\bibfield{author}{\bibinfo{person}{Google}.} \bibinfo{year}{2022}\natexlab{c}.
\newblock \bibinfo{title}{BoostedTreesCreateQuantileStreamResource}.
\newblock
  \bibinfo{howpublished}{\url{https://www.tensorflow.org/api_docs/python/tf/raw_ops/BoostedTreesCreateQuantileStreamResource}}.
\newblock


\bibitem[Google(2022d)]%
        {b43}
\bibfield{author}{\bibinfo{person}{Google}.} \bibinfo{year}{2022}\natexlab{d}.
\newblock \bibinfo{title}{Building the Future of TensorFlow}.
\newblock
  \bibinfo{howpublished}{\url{https://blog.tensorflow.org/2022/10/building-the-future-of-tensorflow.html}}.
\newblock


\bibitem[Han and Cha(2017)]%
        {b48}
\bibfield{author}{\bibinfo{person}{HyungSeok Han} {and}
  \bibinfo{person}{Sang~Kil Cha}.} \bibinfo{year}{2017}\natexlab{}.
\newblock \showarticletitle{Imf: Inferred model-based fuzzer}. In
  \bibinfo{booktitle}{\emph{Proceedings of the 2017 ACM SIGSAC Conference on
  Computer and Communications Security}}. \bibinfo{pages}{2345--2358}.
\newblock


\bibitem[Lattner and Adve(2004)]%
        {b25}
\bibfield{author}{\bibinfo{person}{Chris Lattner} {and} \bibinfo{person}{Vikram
  Adve}.} \bibinfo{year}{2004}\natexlab{}.
\newblock \showarticletitle{LLVM: A compilation framework for lifelong program
  analysis \& transformation}. In \bibinfo{booktitle}{\emph{International
  Symposium on Code Generation and Optimization, 2004. CGO 2004.}} IEEE,
  \bibinfo{pages}{75--86}.
\newblock


\bibitem[LLVM(2022a)]%
        {b26}
\bibfield{author}{\bibinfo{person}{LLVM}.} \bibinfo{year}{2022}\natexlab{a}.
\newblock \bibinfo{title}{Clang}.
\newblock \bibinfo{howpublished}{\url{https://clang.llvm.org}}.
\newblock


\bibitem[LLVM(2022b)]%
        {b34}
\bibfield{author}{\bibinfo{person}{LLVM}.} \bibinfo{year}{2022}\natexlab{b}.
\newblock \bibinfo{title}{libfuzzer}.
\newblock \bibinfo{howpublished}{\url{https://llvm.org/docs/LibFuzzer.html}}.
\newblock


\bibitem[Luo et~al\mbox{.}(2021)]%
        {b15}
\bibfield{author}{\bibinfo{person}{Weisi Luo}, \bibinfo{person}{Dong Chai},
  \bibinfo{person}{Xiaoyue Ruan}, \bibinfo{person}{Jiang Wang},
  \bibinfo{person}{Chunrong Fang}, {and} \bibinfo{person}{Zhenyu Chen}.}
  \bibinfo{year}{2021}\natexlab{}.
\newblock \showarticletitle{Graph-Based Fuzz Testing for Deep Learning
  Inference Engines}. In \bibinfo{booktitle}{\emph{2021 IEEE/ACM 43rd
  International Conference on Software Engineering (ICSE)}}.
  \bibinfo{pages}{288--299}.
\newblock
\urldef\tempurl%
\url{https://doi.org/10.1109/ICSE43902.2021.00037}
\showDOI{\tempurl}


\bibitem[PeachTech(2022)]%
        {b19}
\bibfield{author}{\bibinfo{person}{PeachTech}.}
  \bibinfo{year}{2022}\natexlab{}.
\newblock \bibinfo{title}{peach}.
\newblock \bibinfo{howpublished}{\url{https://www.peach.tech/}}.
\newblock


\bibitem[SWIG(2022)]%
        {b35}
\bibfield{author}{\bibinfo{person}{SWIG}.} \bibinfo{year}{2022}\natexlab{}.
\newblock \bibinfo{title}{SWIG}.
\newblock \bibinfo{howpublished}{\url{https://github.com/swig/swig/}}.
\newblock


\bibitem[Tan et~al\mbox{.}(2022)]%
        {b10}
\bibfield{author}{\bibinfo{person}{Xin Tan}, \bibinfo{person}{Kai Gao},
  \bibinfo{person}{Minghui Zhou}, {and} \bibinfo{person}{Li Zhang}.}
  \bibinfo{year}{2022}\natexlab{}.
\newblock \showarticletitle{An exploratory study of deep learning supply
  chain}. In \bibinfo{booktitle}{\emph{Proceedings of the 44th International
  Conference on Software Engineering}}. \bibinfo{pages}{86--98}.
\newblock


\bibitem[Wang et~al\mbox{.}(2021)]%
        {b16}
\bibfield{author}{\bibinfo{person}{Song Wang}, \bibinfo{person}{Nishtha
  Shrestha}, \bibinfo{person}{Abarna~Kucheri Subburaman},
  \bibinfo{person}{Junjie Wang}, \bibinfo{person}{Moshi Wei}, {and}
  \bibinfo{person}{Nachiappan Nagappan}.} \bibinfo{year}{2021}\natexlab{}.
\newblock \showarticletitle{Automatic Unit Test Generation for Machine Learning
  Libraries: How Far Are We?}. In \bibinfo{booktitle}{\emph{Proceedings of the
  43rd International Conference on Software Engineering}} (Madrid, Spain)
  \emph{(\bibinfo{series}{ICSE '21})}. \bibinfo{publisher}{IEEE Press},
  \bibinfo{pages}{1548–1560}.
\newblock
\showISBNx{9781450390859}
\urldef\tempurl%
\url{https://doi.org/10.1109/ICSE43902.2021.00138}
\showDOI{\tempurl}


\bibitem[Wei et~al\mbox{.}(2022)]%
        {b12}
\bibfield{author}{\bibinfo{person}{Anjiang Wei}, \bibinfo{person}{Yinlin Deng},
  \bibinfo{person}{Chenyuan Yang}, {and} \bibinfo{person}{Lingming Zhang}.}
  \bibinfo{year}{2022}\natexlab{}.
\newblock \showarticletitle{Free lunch for testing: Fuzzing deep-learning
  libraries from open source}.
\newblock \bibinfo{journal}{\emph{arXiv preprint arXiv:2201.06589}}
  (\bibinfo{year}{2022}).
\newblock


\bibitem[Xiao et~al\mbox{.}(2018)]%
        {b9}
\bibfield{author}{\bibinfo{person}{Qixue Xiao}, \bibinfo{person}{Kang Li},
  \bibinfo{person}{Deyue Zhang}, {and} \bibinfo{person}{Weilin Xu}.}
  \bibinfo{year}{2018}\natexlab{}.
\newblock \showarticletitle{Security risks in deep learning implementations}.
  In \bibinfo{booktitle}{\emph{2018 IEEE Security and privacy workshops
  (SPW)}}. IEEE, \bibinfo{pages}{123--128}.
\newblock


\bibitem[Xie et~al\mbox{.}(2022)]%
        {b11}
\bibfield{author}{\bibinfo{person}{Danning Xie}, \bibinfo{person}{Yitong Li},
  \bibinfo{person}{Mijung Kim}, \bibinfo{person}{Hung~Viet Pham},
  \bibinfo{person}{Lin Tan}, \bibinfo{person}{Xiangyu Zhang}, {and}
  \bibinfo{person}{Michael~W Godfrey}.} \bibinfo{year}{2022}\natexlab{}.
\newblock \showarticletitle{Docter: Documentation-guided fuzzing for testing
  deep learning api functions}. In \bibinfo{booktitle}{\emph{Proceedings of the
  31st ACM SIGSOFT International Symposium on Software Testing and Analysis}}.
  \bibinfo{pages}{176--188}.
\newblock


\bibitem[Xie et~al\mbox{.}(2019)]%
        {b14}
\bibfield{author}{\bibinfo{person}{Xiaofei Xie}, \bibinfo{person}{Lei Ma},
  \bibinfo{person}{Felix Juefei-Xu}, \bibinfo{person}{Minhui Xue},
  \bibinfo{person}{Hongxu Chen}, \bibinfo{person}{Yang Liu},
  \bibinfo{person}{Jianjun Zhao}, \bibinfo{person}{Bo Li},
  \bibinfo{person}{Jianxiong Yin}, {and} \bibinfo{person}{Simon See}.}
  \bibinfo{year}{2019}\natexlab{}.
\newblock \showarticletitle{DeepHunter: A Coverage-Guided Fuzz Testing
  Framework for Deep Neural Networks}. In \bibinfo{booktitle}{\emph{Proceedings
  of the 28th ACM SIGSOFT International Symposium on Software Testing and
  Analysis}} (Beijing, China) \emph{(\bibinfo{series}{ISSTA 2019})}.
  \bibinfo{publisher}{Association for Computing Machinery},
  \bibinfo{address}{New York, NY, USA}, \bibinfo{pages}{146–157}.
\newblock
\showISBNx{9781450362245}
\urldef\tempurl%
\url{https://doi.org/10.1145/3293882.3330579}
\showDOI{\tempurl}


\bibitem[Zhang et~al\mbox{.}(2021)]%
        {b23}
\bibfield{author}{\bibinfo{person}{Cen Zhang}, \bibinfo{person}{Xingwei Lin},
  \bibinfo{person}{Yuekang Li}, \bibinfo{person}{Yinxing Xue},
  \bibinfo{person}{Jundong Xie}, \bibinfo{person}{Hongxu Chen},
  \bibinfo{person}{Xinlei Ying}, \bibinfo{person}{Jiashui Wang}, {and}
  \bibinfo{person}{Yang Liu}.} \bibinfo{year}{2021}\natexlab{}.
\newblock \showarticletitle{APICraft: Fuzz Driver Generation for Closed-source
  {SDK} Libraries}. In \bibinfo{booktitle}{\emph{30th {USENIX} Security
  Symposium, {USENIX} Security 2021, August 11-13, 2021}},
  \bibfield{editor}{\bibinfo{person}{Michael Bailey} {and}
  \bibinfo{person}{Rachel Greenstadt}} (Eds.). \bibinfo{publisher}{{USENIX}
  Association}, \bibinfo{pages}{2811--2828}.
\newblock
\urldef\tempurl%
\url{https://www.usenix.org/conference/usenixsecurity21/presentation/zhang-cen}
\showURL{%
\tempurl}


\end{thebibliography}

\end{document}